\pdfoutput=1
\documentclass[12pt]{iopart}
\usepackage{cite,amssymb,amsfonts,graphicx,bm,dcolumn,mathrsfs}
\usepackage{color}
\usepackage{ulem}

\newcommand{\dif}{d}

\newcommand{\kB}{k_B}

\newcommand{\vect}[1]{\bm{#1}}

\newcommand{\bea}{\begin{eqnarray}}
\newcommand{\eea}{\end{eqnarray}}
\newcommand{\be}{\begin{equation}}
\newcommand{\ee}{\end{equation}}
\newcommand{\dd}{\mbox{d}}
\usepackage{epsfig}

\begin{document}

\title{Concavity, Response Functions and Replica Energy}

\author{ Alessandro Campa$^1$, Lapo Casetti$^{2,3}$, Ivan Latella$^4$, Agust\'in P\'erez-Madrid$^5$ and Stefano Ruffo$^6$}
\address{$^1$ National Center for Radiation Protection and Computational Physics, Istituto Superiore di Sanit\`{a},
Viale Regina Elena 299, 00161 Roma, Italy}
\address{$^2$ Dipartimento di Fisica e Astronomia and CSDC, Universit\`a di Firenze,\\and INFN, Sezione di Firenze,
via G. Sansone 1, 50019 Sesto Fiorentino, Italy}
\address{$^3$ INAF-Osservatorio Astrofisico di Arcetri, Largo E. Fermi 5, 50125 Firenze, Italy}
\address{$^4$ Department of Mechanical Engineering, Universit\'e de Sherbrooke, Sherbrooke, Qu\'ebec, J1K 2R1, Canada} 
\address{$^5$ Departament de F\'{i}sica de la Mat\`{e}ria Condensada, Facultat de F\'{i}sica, Universitat de Barcelona, Mart\'{i} i
Franqu\`{e}s 1, 08028 Barcelona, Spain}
\address{$^6$ SISSA, INFN and ISC-CNR, via Bonomea 265, 34136 Trieste, Italy}
\ead{\mailto{alessandro.campa@iss.it}, \mailto{lapo.casetti@unifi.it}, \mailto{ivan.latella@usherbrooke.ca},
\mailto{agustiperezmadrid@ub.edu}
and \mailto{ruffo@sissa.it}}

\begin{abstract}
In nonadditive systems, like small systems or like long-range interacting systems even in the thermodynamic limit,
ensemble inequivalence can be related to the occurrence of negative response functions, this in turn being connected with anomalous
concavity properties of the thermodynamic potentials associated to the various ensembles. We show how the type and number
of negative response functions depend on which of the quantities $E$, $V$  and $N$ (energy, volume and number of particles)
are constrained in the ensemble. In particular, we consider the unconstrained ensemble in which $E$, $V$ and $N$ fluctuate, physically meaningful only for
nonadditive systems. In fact, its partition function is associated to the replica energy, a thermodynamic function that
identically vanishes when additivity holds, but that contains relevant information in nonadditive systems.
\end{abstract}

\noindent{\it Keywords\/}: long-range interactions; non-additive systems; ensemble inequivalence

\maketitle

\section{Introduction}

Additivity can be defined in very simple terms for physical systems. In fact, a system is said to be additive if,
thought as the union of several parts, the energy of interaction between the parts is negligible with respect
to the total energy~\cite{Campa_2014}. In other words, the total energy is the sum of the energies of the different parts, i.e., the
energy is additive. In additive systems, all the extensive quantities, in particular all the thermodynamic potentials and not
only the energy, are additive, and as a consequence these quantities are linear homogeneous functions of the system size;
more precisely, the thermodynamic potentials can be expressed as functions of the intensive variables multiplied by
an extensive variable related to the system size, like the number of constituents $N$ or the volume $V$.

Small systems~\cite{Hill_1963,Hill_1998,Hill_2001,Hill_2002} are obvious examples of nonadditive systems, i.e., systems composed of a number of constituents which is not very large. However, a very important class of nonadditive systems is represented by systems with long-range
interactions including, for instance, self-gravitating systems~\cite{Antonov_1962,Lynden-Bell_1968,
Thirring_1970,Padmanabhan_1990,Lynden-Bell_1999,Chavanis_2002,Chavanis_2006,Latella_2015_b}, plasmas~\cite{Kiessling_2003,Nicholson_1992}, or
fluid dynamics~\cite{Chavanis_2002_b,Robert_1991}.
At variance with small systems with short-range interactions, that become additive by increasing
the number of constituents, systems with long-range interactions are never additive, independently from their size.
It is a simple matter to see that a necessary condition for additivity in macroscopic systems is that the interaction between the constituents decays more rapidly that the inverse of the $d$-th power of the distance, where $d$ is the dimension of the space where the system is embedded.

In nonadditive systems, the thermodynamic potentials are no more linear homogeneous functions of extensive variables; however,
this does not prevent the application of the formalism and of the computational tools of thermodynamics and statistical
mechanics. The statistical mechanics formulation requires proper generalizations~\cite{Campa_2014} to take into account the
nonnegligible interaction between parts of the system, but also a purely thermodynamic description, stemming from the formalism
introduced by Hill for small systems~\cite{Hill_1963}, is possible~\cite{Latella_2015}.

A relevant physical peculiarity of nonadditive systems is that their possible equilibrium states depend on which
thermodynamic quantities are held fixed. These fixed quantities are usually called control parameters (e.g., the total energy $E$
is a control parameter in an isolated system, while the temperature $T$ is a control parameter in a system kept in contact
with a heat bath at that fixed temperature); the other thermodynamic quantities fluctuate around their equilibrium values.
Using the example of the total energy and the temperature, in an additive system the following holds at equilibrium:
fixing the total energy to a value $E$ and finding that the expected value of the temperature is $T_*$, we know that fixing the temperature at  $T=T_*$ will make the expected value $E_*$ of the energy equal to $E$
\footnote{This holds in the so-called thermodynamic limit in which $N$
and $V$ tend to infinite.}. In other words, the equilibrium states do not depend on which control
parameters we use to define them. In the statistical mechanics formalism, this is expressed by the equivalence of the ensembles.
This equivalence is in general absent in nonadditive systems, and physically this implies that there are equilibrium states
defined by given control parameters that do not exist if one chooses another set of control parameters. This will be
stated in more precise terms later.

Ensemble inequivalence is related to the occurrence of negative response functions, this in turn being related to anomalous
concavity properties of the thermodynamic potentials associated to the various ensembles. In this paper we focus exactly on this issue,
making a survey of all the statistical ensembles and thermodynamic potentials, showing in each case the response function that can have
a negative value. We stress that ensemble inequivalence can give rise to negative response functions, but that this is not necessary,
since there can be inequivalence without negative response functions. We will be more precise about this issue in the following. For the moment we underline that a negative response function implies ensemble inequivalence, while ensemble inequivalence does not necessarily imply a
negative response function. 

The paper is organized as follows. In Section~\ref{therm_repl_ens} we show how the replica energy can be introduced, a relevant thermodynamic function for nonadditive systems, and present the associated statistical ensemble that is treated in more details later.
In Sections \ref{ensemble_inequivalence} and \ref{from_micr_to_uncon}, we describe the relation between ensemble inequivalence and the anomalous
concavity properties and response functions; the latter section is dedicated to the ensemble associated to the replica energy, while the former
section concerns the other ensembles. In the last section we present a discussion with concluding remarks.

\section{Thermodynamics, replica energy, statistical ensembles}
\label{therm_repl_ens}

To obtain the thermodynamic properties of a system from the principles of statistical mechanics, one considers a great
number $\mathscr{N}$ of independent replicas of the system, namely, a statistical ensemble. The replicas are identical in nature, but they differ
in phase, that is, in their condition with respect to configuration and velocity~\cite{Gibbs}.
If the energy, entropy, volume and number of particles of the system under consideration are $E$, $S$, $V$ and $N$, respectively,
the corresponding quantities of the ensemble are $E_{\rm t}=\mathscr{N}E$,
$S_{\rm t}=\mathscr{N}S$, $V_{\rm t}=\mathscr{N}V$ and $N_{{\rm t}}=\mathscr{N}N$. Energy variations in the ensemble
satisfy the general thermodynamic relation~\cite{Hill_1963}
\begin{equation}
\dd E_{\rm t}=T\dd S_{\rm t}-P\dd V_{\rm t}+\mu\dd N_{{\rm t}}+ \mathscr{E}\dd \mathscr{N},
\label{Hill_equation}
\end{equation}
where $T$ is the temperature, $P$ is the pressure exerted on the boundary of the systems,
and $\mu$ is the chemical potential of a single system. The last term on the right-hand side of equation~(\ref{Hill_equation})
accounts for the energy variation when the number of members of the ensemble $\mathscr{N}$ varies, holding $S_{\rm t}$,
$V_{\rm t}$ and $N_{{\rm t}}$ constant.
The replica energy $\mathscr{E}$, formally given by
\begin{equation}
\mathscr{E}=\left(\frac{\partial E_{\rm t}}{\partial\mathscr{N}}\right)_{S_{\rm t}, V_{\rm t},N_{{\rm t}}},
\end{equation}
vanishes if the system is additive; this can be derived using the fact that for additive systems the extensive variables are linear homogeneous
functions of the system size~\cite{Latella_2015}. In the case in which all single systems properties are held constant, equation~(\ref{Hill_equation})
can be written
$E\dd \mathscr{N}=TS\dd \mathscr{N}-PV\dd \mathscr{N}+\mu N\dd \mathscr{N}+ \mathscr{E}\dd \mathscr{N}$,
which can be integrated from 0 to $\mathscr{N}$ to give
$E_{\rm t}=TS_{\rm t}-P V_{\rm t}+\mu N_{{\rm t}}+ \mathscr{E}\mathscr{N}$.
Dividing by $\mathscr{N}$ the latter equation gives 
\begin{equation}
E=TS-P V+\mu N+ \mathscr{E}, 
\label{Euler_relation}
\end{equation}
which relates the properties of a single system with the replica energy. By differentiation one obtains:
\begin{equation}
\dd \mathscr{E} = \dd E -T\dif S -S \dd T +P\dd V + V\dd P -\mu \dd N -N\dd \mu \, .
\label{generalized_prov}
\end{equation}
Thus, one can exploit the first law of thermodynamics, expressed by
\begin{equation}
\dd E = T\dd S-P\dd V+\mu\dd N \, ,
\label{dif_energy}
\end{equation}
to obtain
\begin{equation}
\dd \mathscr{E} =-S\dd T+V\dd P-N\dd \mu \, .
\label{generalized-GD}
\end{equation}
The above equation generalizes the usual Gibbs-Duhem equation for additive systems, which is obtained by setting to zero the left hand side of Eq.~(\ref{generalized-GD}). Since, in general, the usual Gibbs-Duhem equation does not hold for nonadditive systems~\cite{Latella_2013}, there exists the possibility of taking $T$, $P$ and $\mu$ as independent variables, a fact that is forbidden when $\mathscr{E}=0$. Moreover, as can be seen from Eq.~(\ref{Euler_relation}), we highlight that when $\mathscr{E}\neq0$, the Gibbs free energy $G=E-TS+PV$ is not equal to $\mu N$.

Depending on the control parameters defining the state of the system, certain quantities fluctuate and other
quantities are fixed. Distinguishing between these two kinds of quantities is relevant here, and it is convenient to set now the
notation that will be used to indicate such a distinction when necessary: if the energy $E$, volume $V$, or number of
particles $N$ are not control parameters, they are fluctuating quantities and will be denoted with a bar by $\bar{E}$, $\bar{V}$,
and $\bar{N}$, respectively. Equations (\ref{Euler_relation}), (\ref{dif_energy}), and (\ref{generalized-GD}) are general
relations at a thermodynamic level and have to be understood for quantities with or without bars.

We will refer
to the variables $E$, $V$, and $N$ as constraint variables, and ensembles in which at least one of the constraint variables
is a control parameter will be termed as constrained ensembles. The thermodynamic properties of an isolated system
are obtained from a completely constrained ensemble in which all the constraint variables are control parameters; as well known,
this is the microcanonical ensemble. On the other hand, if none of the constraint variables is a control parameter, the system is
said to be completely open and the associated ensemble is the unconstrained ensemble.

We have reminded above that ensemble inequivalence is associated to the fact that, for nonadditive systems,
the possible equilibrium configurations depend on the specific control parameters used to define its state.
Therefore, the thermodynamics of the system must be necessarily derived from the characteristic function (the entropy or the
free energies) in the ensemble associated to the particular set of control parameters under consideration. In doing so, it is
in general possible to obtain the replica energy from the corresponding characteristic function, except in the case where the
replica energy itself is the characteristic function corresponding to a particular set of control parameters. In fact, the
replica energy is the free energy associated to the unconstrained ensemble where the corresponding control parameters are
$T$, $P$ and $\mu$~\cite{Latella_2017}, namely,
\begin{equation}
\mathscr{E}(T,P,\mu)\equiv- \kB T\ln\Upsilon(T,P,\mu),
\end{equation}
where
\begin{equation}
\Upsilon(T,P,\mu)= \int\dd E\int\dd V\sum_{N=0}^\infty \omega(E,V,N) \e^{-(E + PV -\mu N )/(\kB T)}.
\label{unconstrained_partition_function}
\end{equation}
is the unconstrained partition function, $\omega(E,V,N)$ being the microcanonical density of states, which is defined below, and $\kB$ the Boltzmann constant.
In Section \ref{from_micr_to_uncon} we will come back to the relation between the microcanonical density of state (and the associated
microcanonical entropy) and the replica energy.

\section{Response functions and ensemble inequivalence}
\label{ensemble_inequivalence}

In this and in the next section, we analyze the relation between ensemble inequivalence and the occurrence
of negative response functions. Ensemble inequivalence can be studied with the help of the properties of the Legendre-Fenchel
transformation; this approach, already well documented for constrained ensembles~\cite{Ellis_2000,Touchette_2004,Touchette_2009},
can be extended to the case of the unconstrained ensemble. It is the Legendre-Fenchel transformation that allows one to connect ensemble
inequivalence and negative response functions. In this paper we are  
particularly interested in the inequivalence between the unconstrained ensemble and the other ensembles. However, it is
instructive to consider first the inequivalence between constrained ensembles; this will be done in this section,
showing the associated anomalous response functions. 
The unconstrained ensemble will be considered in Section~\ref{from_micr_to_uncon}.
In the following, we use units in which the Boltzmann constant $\kB$ is set to unity.

To begin we consider the microcanonical and canonical ensembles for a system described by the Hamiltonian $\mathcal{H}(\vect{p},\vect{q})$
with $\vect{p}=(\vect{p}_1,\ldots,\vect{p}_N)$ and $\vect{q}=(\vect{q}_1,\ldots,\vect{q}_N)$, where $\vect{p}_i\in\mathbb{R}^d$
and $\vect{q}_i\in\mathbb{R}^d$ are the momentum and position of particle $i$, respectively, and $d$ is the dimensionality of
the system. Just for completeness, we remind that, while the constraint variables $E$, $V$ and $N$ are the control parameters of the
microcanonical ensemble, the control parameters of the canonical ensemble are $T$, $V$  and $N$. The microcanonical density of states is given by
\begin{equation}
\omega(E,V,N)=\frac{1}{h^{dN}N!}\int\delta(E-\mathcal{H}(\vect{p},\vect{q}))\, \dd^{2dN}\vect{\Gamma},
\label{density_of_states_microcanonical}
\end{equation}
where $h$ is a constant and $\dd^{2dN}\vect{\Gamma}=\prod_{i=1}^N\dd^d\vect{p}_i\dd^d\vect{q}_i$, while the canonical partition
function is
\begin{equation}
Z\left(T,V,N\right)= \frac{1}{h^{dN}N!} \int \e^{-\mathcal{H}(\vect{p},\vect{q})/T}\, \dd^{2dN}\vect{\Gamma}.
\label{partition_function_canonical_definition}
\end{equation}
Taking advantage of the Dirac $\delta$ in (\ref{density_of_states_microcanonical}), after posing $Z=\e^{-\mathcal{F}}$ we rewrite the
canonical partition function as
\begin{equation}
\e^{-\mathcal{F}(\beta,V,N)}=\int\dd E\, \omega(E,V,N)\,\e^{-\beta E}=\int\dd E\,\e^{S(E,V,N)- \beta E},
\label{partition_function_canonical_definition_2}
\end{equation}
with $\beta = 1/T$ being the inverse canonical temperature, $\mathcal{F}=\beta F$ the rescaled Helmholtz free energy, and $S=S(E,V,N)$,
the logarithm of the density of states $\omega$, the microcanonical entropy. In the large $N$ limit, we can compute the integral on the right-hand
side of (\ref{partition_function_canonical_definition_2}) using the saddle-point approximation and write
\begin{equation}
\mathcal{F}(\beta,V,N)=\inf_E\left[\beta E-S(E,V,N)\right].
\label{rescaled_F_LFT}
\end{equation}
We thus obtain the rescaled Helmholtz free energy as the Legendre-Fenchel transform of the microcanonical entropy with respect to
the energy~\cite{Ellis_2000,Touchette_2004,Touchette_2009,Campa_2014}, which reduces to the usual Legendre transformation if the
entropy is differentiable and concave in $E$ at constant $V$ and $N$.

On the one hand, the Legendre-Fenchel transformation of any function, as defined in (\ref{rescaled_F_LFT}), is always a globally concave
function~\cite{Touchette_2009}. This very remarkable property guarantees that the rescaled free energy $\mathcal{F}$ is always globally
concave with respect to $\beta$. For convenience we recall in Appendix 1 the definition of locally and globally concave (and convex) functions,
of concave (and convex) envelope, together with some properties of the Legendre-Fenchel transformation and its
relations with concave functions. In the following we will refer to these definitions and relations several times, therefore the reader not familiar
with them should read Appendix 1 at this point. Using that 
\begin{equation}
\bar{E} =\left(\frac{\partial \mathcal{F}}{\partial \beta}\right)_{V,N},
\label{mean_energy_canonical} 
\end{equation}
the concavity of $\mathcal{F}$ with respect to $\beta$ means that
\begin{equation}
\left(\frac{\partial\bar{E}}{\partial\beta}\right)_{V,N}= \left(\frac{\partial^2\mathcal{F}}{\partial\beta^2}\right)_{V,N}\leq0,
\end{equation}
which ensures that the response function 
\begin{equation}
C_{V,N}=\left(\frac{\partial\bar{E}}{\partial T}\right)_{V,N}\geq 0,
\label{C_V_canonical}
\end{equation}
that is, the heat capacity, is a nonnegative quantity in the canonical ensemble. This statement is valid regardless of the differentiability
of $\mathcal{F}$. If $\mathcal{F}$ is twice-differentiable, then $C_{V,N}$ is continuous, otherwise it has discontinuities, or it can even
diverge, if $\mathcal{F}$ is not differentiable, for the values of $\beta$ where the derivative of this function is not continuous; however, the inequality in Eq. (\ref{C_V_canonical}) is always satisfied in the
canonical ensemble, since $\mathcal{F}$ is always globally concave. For additive systems, the heat capacity
is a nonnegative quantity also in the microcanonical ensemble (furthermore, it coincides with that in the canonical ensemble), since for
these systems it can be proved that the microcanonical entropy $S(E,V,N)$ is globally concave with respect to $E$~\cite{ruelle} and, as
remarked in Appendix 1, globally concave functions coincide with their concave envelope. Actually, it can be proved
that for additive systems $S(E,V,N)$ is globally concave also with respect to $V$, and in addition it is globally completely concave in
the $(E,V)$ plane~\cite{ruelle}.  
On the other hand, the lack of additivity can induce the lack of global concavity in the microcanonical entropy as a
function of the energy. Hence, the quantity
\begin{equation}
\frac{1}{C_{V,N}}=\left(\frac{\partial T}{\partial E}\right)_{V,N}=-T^2\left(\frac{\partial^2 S}{\partial E^2}\right)_{V,N}
\label{C_V_microcanonical}
\end{equation}
can be negative in the microcanonical ensemble (hereafter we use, for simplicity, the same symbol to represent the response
functions in the different ensembles). Referring to Figure~\ref{fig_concave} in Appendix 1, we have a negative microcanonical heat capacity $C_{V,N}$
for a range of $E$ values if the entropy has a behavior similar to that of the upper curve or the middle curve; if the behavior is similar to that of the lower
curve the heat capacity is positive except for the $E$ value where the cusp occurs, and where it is not defined. Note that the middle curve presents both
features shown separately by the other two curves: it has a range of $E$ where $C_{V,N}$ is negative and also a point of discontinuity. In all these cases the
microcanonical entropy
does not coincide with its concave envelope; its Legendre-Fenchel transform, i.e., the function $\mathcal{F}(\beta,V,N)$, will have at least a
point $\beta$, for the given $V$ and $N$ values, where its first derivative with respect to $\beta$ is not defined (see Appendix 1). Thus, also
the associated response function, the heat capacity, is not defined there. This point marks
the occurrence of a first order phase transition in the canonical ensemble. We remark that, apart from such points, the canonical
heat capacity (\ref{C_V_canonical}) is perfectly defined and always positive.
Negative heat capacities in the microcanonical ensemble can occur since they are not forbidden by any fundamental requirement.
Besides, according to equation~(\ref{C_V_canonical}), equilibrium states with negative heat capacity cannot be realized if the
system is put in contact with an infinite thermal bath (canonical ensemble).
It is therefore clear that states associated to energy values where the entropy does not coincide with its concave envelope have no correspondence in the canonical ensemble.

To summarize the main result, if the microcanonical entropy does not coincide with its
concave envelope with respect to $E$, the microcanonical and canonical ensembles are not
equivalent~\cite{Bouchet_2005,Ellis_2000,Touchette_2004,Touchette_2009,Campa_2014}. In this case, the function $\mathcal{F}$ presents
at least a point of discontinuous derivative with respect to $\beta$, associated to a first order phase transition\footnote{ The function $\mathcal{F}$ has a discontinuous derivative with respect to $\beta$ also in the limiting case where the microcanonical entropy does coincide with its concave envelope, but the latter is a linear function of the energy in a given interval. This case has been referred to as ``partial equivalence''~\cite{Lapo} because there is equivalence but not one-to-one: a single value of $\beta$ corresponds to a whole interval of values of the energy. This may happen also in additive systems and indeed it happens whenever the system undergoes a discontinuous phase transition, e.g., when there is a change of state like a liquid-gas phase transition.}.

It is useful to stress the physical reason that permits having a negative heat capacity in the microcanonical ensemble,
while this is not allowed in the canonical ensemble. In the microcanonical ensemble, the energy is fixed, and it can be given a value belonging
to the energy range of convexity. In the canonical ensemble the energy can fluctuate, and it can be easily seen that if a system at a given energy
$E$ in the energy range of convexity, with expected value of the temperature equal to $T_*$, is put in contact with a heat bath at temperature $T=T_*$, it is unstable with respect to energy fluctuations, and it will acquire an expected value of the energy where the
associated temperature is also $T_*$, but that it is located in an energy range of concavity of the microcanonical entropy.
A state with an energy where the microcanonical entropy is locally concave, but that does not belong to the range where it coincides with
its concave envelope, is metastable when put in contact with a heat bath at the corresponding temperature $T_*$, i.e., it is stable
with respect to sufficiently small energy fluctuations, but not with respect to general fluctuations; namely it is not globally stable
and then it cannot be defined as an equilibrium state~\cite{Campa_2014}.

In the microcanonical ensemble, nonadditive systems could exhibit a convex region  in the entropy as a function of the other
constraint variables, $V$ or $N$, or, more generally, ranges where the entropy does not coincide
with its concave envelope with respect to one or both of these variables (while for additive systems the entropy is globally concave with respect to $V$ and $N$). For those variables, such
anomalus behavior is inherited by the canonical ensemble, since in this case both $V$
and $N$ are control parameters as well, and the Legendre-Fenchel transformation does not involve them. However, we point out the following.
While ``normal'' behavior of the microcanonical entropy $S(E,V,N)$ is represented by global concavity with respect to $V$ and $N$, ``normal'' behavior
of the rescaled Helmholtz free energy $\mathcal{F}(\beta,V,N)$ [or of the free energy $F(\beta,V,N)$] is represented by global convexity with respect to
these variables,
since in the Legendre-Fenchel transform (\ref{rescaled_F_LFT}) the microcanonical entropy appears with the minus sign. Thus, in nonadditive systems
where the microcanonical entropy can have ranges of convexity in $V$ and/or in $N$, correspondingly the Helmholtz free energy will have
ranges of concavity in $V$ and/or in $N$.

Let us now turn to the grand canonical ensemble. In this ensemble, in addition to the energy also the number of particles is not constrained.
The control parameters of this ensemble are $\mu$, $T$ and $V$.
The grand canonical partition function $\Xi=\e^{-\mathcal{L}}$ can be written as
\begin{equation}
\e^{-\mathcal{L}(\alpha,\beta,V)}= \sum_{N=0}^\infty \e^{\mu N/T}Z(T,V,N)=\sum_{N=0}^\infty \e^{-\alpha N-\mathcal{F}(\beta,V,N)},
\label{partition_function_grand_canonical_definition_2}
\end{equation}
where $\alpha=-\mu/T$.
The rescaled grand potential $\mathcal{L}=\beta\Omega$ is thus given by the term that dominates the sum according to
\begin{equation}
\mathcal{L}(\alpha,\beta,V)=\inf_N\left[\alpha N+\mathcal{F}(\beta,V,N)\right] \, ,
\label{rescaled_Omega_LFT}
\end{equation}
which is the Legendre-Fenchel transform of $-\mathcal{F}=-\beta F$ with respect to $N$. This expression (\ref{rescaled_Omega_LFT}) assures that
$\mathcal{L}(\alpha,\beta,,V)$ is always globally concave in $\alpha$, and that its concavity with respect to $\beta$ is inherited
from that of $\mathcal{F}$. Using Eq. (\ref{rescaled_F_LFT}) we can also write
\begin{equation}
\mathcal{L}(\alpha,\beta,V)=\inf_{E,N}\left[\alpha N+\beta E-S(E,V,N)\right] \, .
\label{rescaled_Omega_LFT_2}
\end{equation}
From this expression we infer that, in addition, $\mathcal{L}(\alpha,\beta,,V)$ is globally completely concave in the plane $(\alpha,\beta)$.
Thus, from
\begin{equation}
\bar{N}=\left(\frac{\partial \mathcal{L}}{\partial\alpha}\right)_{\beta,V} \, ,
\end{equation}
we have
\begin{equation}
\left(\frac{\partial\bar{N}}{\partial\alpha}\right)_{\beta,V}=
\left(\frac{\partial^2\mathcal{L}}{\partial\alpha^2}\right)_{\beta,V}\leq0 \, ,
\end{equation}
so that in the grand canonical ensemble
\begin{equation}
M_{T,V}\equiv\left(\frac{\partial\bar{N}}{\partial\mu}\right)_{T,V}\geq0.
\label{M_T_V_grand}
\end{equation}
Here $M_{T,V}$ is a response function, just as the heat capacity; it tells us that in the grand canonical ensemble the number
of particles increases whenever the chemical potential increases, holding $T$ and $V$ constant.
We can repeat here the observation made for the canonical heat capacity (\ref{C_V_canonical}). Thus, the positivity of $M_{T,V}$
in the grand canonical ensemble is valid regardless of the differentiability
of $\mathcal{L}$. If $\mathcal{L}$ is twice-differentiable, then $M_{T,V}$ is continuous, otherwise it has discontinuities, or it can even
diverge, if $\mathcal{L}$ is not differentiable, for the values of $\mu$ where the derivative of this function is not continuous; however, the inequality in
Eq. (\ref{M_T_V_grand}) is always satisfied in the grand canonical ensemble since $\mathcal{L}$ is always globally concave.

As a side remark, we note that this response function
can be written as $M_{T,V}=\beta N/\Gamma$, where $\Gamma$ is the thermodynamic factor given by~\cite{Schnell_2012} 
\begin{equation}
\frac{1}{\Gamma}=\frac{1}{\beta}\left(\frac{\partial \ln\bar{N}}{\partial\mu}\right)_{T,V}.
\end{equation}
For macroscopic short-range interacting systems the usual Gibbs-Duhem equation holds, and the function $M_{T,V}$
can be directly related to the isothermal compressibility\footnote{For macroscopic short-range interacting systems, using
$n=\bar{N}/V$ we can write
\begin{equation*}
M_{T,V}=\left(\frac{\partial\bar{N}}{\partial\mu}\right)_{T,V}=
V\left(\frac{\partial n}{\partial\mu}\right)_{T}=
V\left(\frac{\partial P}{\partial\mu}\right)_{T} \left(\frac{\partial n}{\partial P\vphantom{\mu}}\right)_{T}.
\end{equation*}
Since the Gibbs-Duhem holds in this case ($\mathscr{E}=0$), under isothermal conditions we have $\dd P=n\dd \mu$.
Hence, using that $\partial n/\partial P=-n^2\partial(1/n)/\partial P$, $\kappa_T$ is related to $M_{T,V}$ according to
\begin{equation*}
M_{T,V}=Vn\left(\frac{\partial n}{\partial P\vphantom{\mu}}\right)_{T}=
-\frac{\bar{N}n}{V}\left(\frac{\partial V}{\partial P\vphantom{\mu}}\right)_{T,\bar{N}}=\bar{N}n\kappa_T.
\end{equation*}
} $\kappa_T$. In the latter case, the sign of $M_{T,V}$ and that of $\kappa_T$ are the same, namely, they are both
positive quantities.
However, if the replica energy is different from zero, as in nonadditive systems, the signs of these response functions are independent from each other,
in general.

Concerning the issue of ensemble inequivalence, in the canonical ensemble there is no mechanism ensuring that for nonadditive systems
the Helmholtz free energy is convex with respect to $N$. Therefore the quantity
\begin{equation}
\frac{1}{M_{T,V}}=\left(\frac{\partial\mu}{\partial N}\right)_{T,V}= \left(\frac{\partial^2 F}{\partial N^2}\right)_{T,V}
\label{M_T_V_canonical}
\end{equation}
could be negative. Again, in perfect analogy to the relation between microcanonical and canonical ensembles, we have the following. If the rescaled Helmholtz free energy does not coincide with its convex envelope with respect to $N$, then its convex Legendre-Fenchel transform $-\mathcal{L}$
will have at least a point $\alpha$, for the given $V$ and $\beta$ values, where its first derivative with respect to $\alpha$ is not defined,  marking the occurrence of a first order phase transition.
Apart from this isolated point, or points, $M_{T,V}$ is perfectly defined and always positive in the grand canonical ensemble. On the other hand,
the response function $M_{T,V}$ in the canonical ensemble can be negative, if the rescaled Helmholtz free energy has a range where it is not locally convex
with respect to $N$, or could have points where it is not defined, or both; these three cases correspond to the upper, lower and middle curves in Figure~\ref{fig_concave},
respectively.

Summarizing the main result, if the rescaled Helmholtz free
energy does not coincide with its convex envelope with respect to $N$, the canonical and grand canonical ensembles are
not equivalent. In this case, the function $\mathcal{L}$ presents
at least a point of discontinuous derivative with respect to $\alpha$, associated to a first order phase transition. If an equilibrium canonical state in which the rescaled Helmholtz free energy does not coincide with its convex envelope with respect to $N$ is put in contact with a reservoir with its same chemical potential and with which
it can exchange particles, then it becomes either unstable or not globally stable.

We now consider the isothermal-isobaric ensemble, where the volume is not a control parameter; the control parameters
are $N$, $T$ and $P$. The associated partition function
$\Delta=\e^{-\mathcal{G}}$ is written as
\begin{equation}
\e^{-\mathcal{G}(N,\beta,\gamma)}=\int\dif V\, \e^{-PV/T}Z(T,V,N)=\int\dif V\e^{-\gamma V-\mathcal{F}(\beta,V,N)},
\end{equation}
where $\gamma=P/T$.
Hence, the saddle-point approximation gives the rescaled Gibbs free energy $\mathcal{G}=\beta G$ as
\begin{equation}
\mathcal{G}(N,\beta,\gamma)=\inf_V\left[\gamma V+\mathcal{F}(\beta,V,N)\right],
\label{rescaled_G_LFT}
\end{equation}
which is the Legendre-Fenchel transform of $-\mathcal{F}$ with respect to $V$. Moreover, using (\ref{rescaled_F_LFT}) we
can also write
\begin{equation}
\mathcal{G}(N,\beta,\gamma)=\inf_{E,V}\left[\beta E+\gamma V-S(E,V,N)\right],
\label{rescaled_G_LFT_2}
\end{equation}
from which we infer that $\mathcal{G}(N,\beta,\gamma)$ is concave in both $\beta$ and $\gamma$; moreover, it is completely concave in the plane $(\beta,\gamma)$. In particular, using that
\begin{equation}
\bar{V} = \left(\frac{\partial \mathcal{G}}{\partial \gamma}\right)_{N,\beta} \, ,
\label{vabrfromiso}
\end{equation}
we can assert that
\begin{equation}
\left(\frac{\partial\bar{V}}{\partial \gamma}\right)_{N,\beta}=
\left(\frac{\partial^2 \mathcal{G}}{\partial \gamma^2}\right)_{N,\beta}\leq0,
\end{equation}
and therefore that the isothermal compressibility in the isothermal-isobaric ensemble is nonnegative,
\begin{equation}
\kappa_T =-\frac{1}{\bar{V}}\left(\frac{\partial \bar{V}}{\partial P}\right)_{T,N}\geq0 \, .
\end{equation}
This is what we expect on physical grounds, since states with negative $\kappa_T$ cannot be stable under volume fluctuations.
For convenience, instead of the isothermal compressibility $\kappa_T$ we can consider the quantity
$K_{T,N}=\bar{V}\kappa_T $ as a response function, where the subscript $N$ is written to emphasize that it is also computed
at constant number of particles. Then, in the isothermal-isobaric ensemble
\begin{equation}
K_{T,N} =-\left(\frac{\partial \bar{V}}{\partial P}\right)_{T,N}\geq0.
\label{ktnisobar}
\end{equation}
The same argument made before applies. Thus, the positivity of $K_{T,N}$
in the isothermal-isobaric ensemble is valid regardless of the differentiability of $\mathcal{G}$. If $\mathcal{G}$ is twice-differentiable,
then $K_{T,N}$ is continuous, otherwise it has discontinuities, or it can even diverge, if $\mathcal{G}$ is not differentiable, for
the values of $P$ where the derivative of this function is not continuous; however, the inequality in
Eq. (\ref{ktnisobar}) is always satisfied in the isothermal-isobaric ensemble, since $\mathcal{G}$ is always globally concave.
However, in the canonical ensemble the volume is a control parameter, i.e., it is fixed in the equilibrium configuration.
The Helmholtz free energy for nonadditive systems is not necessarily convex with respect to $V$, so that
states with negative isothermal compressibility or, equivalently, negative $K_{T,N}$ can be realized. In fact, in the canonical ensemble we have
\begin{equation}
\frac{1}{K_{T,N}}=-\left(\frac{\partial P}{\partial V}\right)_{T,N}= \left(\frac{\partial^2 F}{\partial V^2}\right)_{T,N},
\end{equation}
which is not restricted to be a positive quantity.
An argument analogous to that already used before implies the following. If the rescaled Helmholtz
free energy does not coincide with its convex envelope with respect to $V$, then its convex Legendre-Fenchel transform $-\mathcal{G}$
will have at least a point $\gamma$, for the given $N$ and $\beta$ values, where its first derivative with respect to $\gamma$ is not defined,
 marking the occurrence of a first order phase transition.
Apart from this isolated point, or points, $K_{T,N}$ is perfectly defined and always positive in the isothermal-isobaric ensemble. On the other hand,
the response function $K_{T,N}$ in the canonical ensemble can be negative at points where the rescaled Helmholtz free energy is not locally convex
with respect to $V$, or could have points where it is not defined, or both (the three cases represented in Figure~\ref{fig_concave}).

According to the previous discussion, we conclude that if the rescaled Helmholtz free
energy does not coincide with its convex envelope with respect to $V$, the canonical and isothermal-isobaric ensembles are
not equivalent. In this case, the function $\mathcal{G}$ presents
at least a point of discontinuous derivative with respect to $\gamma$, associated to a first order phase transition.
If an equilibrium canonical state in which the rescaled Helmholtz free energy does not coincide with its convex envelope with respect to $V$
is put in contact with an environment with its same pressure, then it becomes either unstable or not globally stable.

\section{From microcanonical entropy to rescaled replica energy}
\label{from_micr_to_uncon}

Continuing the discussion of the preceding section, here we focus on the unconstrained ensemble and its connection with
the other ensembles. Since the unconstrained ensemble describes the thermodynamics of completely open systems, it can be
seen as the opposite situation of the one described by the microcanonical ensemble where the systems are isolated. Such an
opposite situation is reflected in the curvature properties of the thermodynamic characteristic functions. We shall
see that the characteristic function of completely open systems, the rescaled replica energy, possesses always a very
well defined concavity with respect to all its natural variables (none of them being a constraint variable), while, as
noted previously, the microcanonical entropy can be nonconcave in any of its natural variables (all of them being
constraint variables).

In the case where the energy, volume, and number of particles fluctuate, from (\ref{unconstrained_partition_function}) we
can write the unconstrained partition function $\Upsilon=\e^{-\mathcal{R}}$ as a function of the microcanonical entropy, that is
\begin{equation}
\e^{-\mathcal{R}(\alpha,\beta,\gamma)}= \int\dif E\int\dif V\sum_{N=0}^\infty \e^{S(E,V,N)-\alpha N- \beta E-\gamma V} \, .
\label{rescaled_replica_micro_entropy}
\end{equation}
Similarly to the other ensembles, the rescaled replica energy $\mathcal{R}$ and the replica energy $\mathscr{E}$ are related
by $\mathcal{R}=\beta\mathscr{E}$. We note that considering the set of
control parameters $\alpha=-\mu/T$, $\beta=1/T$, and $\gamma=P/T$ is completely equivalent to considering $T$, $P$, and $\mu$.
Evaluating (\ref{rescaled_replica_micro_entropy}) in a saddle-point approximation we have
\begin{equation}
\mathcal{R}(\alpha,\beta,\gamma)=\inf_{E,V,N}\left[ \alpha N+ \beta E+\gamma V-S(E,V,N)\right],
\label{rescaled_replica_LFT}
\end{equation}
which ensures that $\mathcal{R}(\alpha,\beta,\gamma)$ is completely concave, implying that it is also separately concave in
$\alpha$, $\beta$, and $\gamma$.
Therefore, in the unconstrained ensemble we get
\begin{eqnarray}
\left(\frac{\partial \bar{N}}{\partial\alpha}\right)_{\beta,\gamma}&=
\left(\frac{\partial^2 \mathcal{R}}{\partial\alpha^2}\right)_{\beta,\gamma}\leq0\label{response_alpha},\\
\left(\frac{\partial \bar{E}}{\partial\beta}\right)_{\alpha,\gamma}&=
\left(\frac{\partial^2 \mathcal{R}}{\partial\beta^2}\right)_{\alpha,\gamma}\leq0\label{response_beta},\\
\left(\frac{\partial \bar{V}}{\partial\gamma}\right)_{\alpha,\beta}&=
\left(\frac{\partial^2 \mathcal{R}}{\partial\gamma^2}\right)_{\alpha,\beta}\leq0\label{response_gamma}.
\end{eqnarray}
Using that $\alpha=-\mu/T$, $\beta=1/T$, and $\gamma=P/T$, equations (\ref{response_alpha}), (\ref{response_beta}), and
(\ref{response_gamma}) imply that the response functions
\begin{eqnarray}
M_{T,P}&\equiv&\left(\frac{\partial \bar{N}}{\partial\mu}\right)_{T,P}\geq0,\label{M_P}\\
C_{\alpha,\gamma}&\equiv&\left(\frac{\partial \bar{E}}{\partial T}\right)_{\mu/T,P/T}\geq0,\label{C_P}\\
K_{T,\mu}&\equiv&-\left(\frac{\partial \bar{V}}{\partial P}\right)_{T,\mu}\geq0,\label{K_mu}
\end{eqnarray}
respectively, are nonnegative in the unconstrained ensemble.
As before, there could be isolated points where these response functions are not defined;
apart from these isolated points, the response functions are perfectly defined and are always positive.

The rescaled replica energy can be related
to the rescaled grand potential via
\begin{equation}
\e^{-\mathcal{R}(\alpha,\beta,\gamma)}= \int\dif V\,\Xi(T,V,\mu)\, \e^{-P V/T}= \int\dif V\, \e^{-\mathcal{L}(\alpha,\beta,V)-\gamma V}.
\label{rescaled_replica_grand_canonical}
\end{equation}
Using the saddle-point approximation, we then have
\begin{equation}
\mathcal{R}(\alpha,\beta,\gamma)=\inf_{V}\left[ \gamma V+ \mathcal{L}(\alpha,\beta,V)\right],
\label{rescaled_replica_LFT_2}
\end{equation}
so that $\mathcal{R}$ is expressed as the Legendre-Fenchel transform of $-\mathcal{L}=-\beta\Omega$ with respect to $V$.
The grand potential is not necessarily a convex function in $V$ for nonadditive systems, so that
the response function $K_{T,\mu}$, given in the grand canonical ensemble by
\begin{equation}
\frac{1}{K_{T,\mu}}=-\left(\frac{\partial P}{\partial V}\right)_{T,\mu}= \left(\frac{\partial^2 \Omega}{\partial V^2}\right)_{T,\mu},
\end{equation}
can be a negative quantity. If the grand potential does not coincide with its convex envelope with respect to $V$, the grand canonical
and unconstrained ensembles are not equivalent. As before,
the response function $K_{T,\mu}$ in the grand canonical ensemble can be negative where the rescaled grand potential is not locally convex
with respect to $V$, or could have points where it is not defined, or both (the three cases of Figure~\ref{fig_concave}).

We observe here that if the rescaled grand potential
energy does not coincide with its convex envelope with respect to $V$, the grand canonical and the unconstrained ensembles are
not equivalent. In this case, the function $\mathcal{R}$ presents
at least a point of discontinuous derivative with respect to $\gamma$, associated to a first order phase transition.
If an equilibrium grand canonical state in which the rescaled grand potential does not coincide with its convex envelope with respect to $V$
is put in contact with an environment with its same pressure, then it becomes either unstable or not globally stable.

Furthermore, we can also write
\begin{equation}
\e^{-\mathcal{R}(\alpha,\beta,\gamma)}=
\sum_{N=0}^\infty \Delta(T,P,N)\,\e^{\mu N/T}= \sum_{N=0}^\infty \e^{-\mathcal{G}(N,\beta,\gamma)-\alpha N},
\label{rescaled_replica_isothermal}
\end{equation}
which relates the rescaled replica energy to the rescaled Gibbs free energy, and therefore we obtain $\mathcal{R}$ as the
Legendre-Fenchel transform of $-\mathcal{G}=-\beta G$ with respect to $N$,
\begin{equation}
\mathcal{R}(\alpha,\beta,\gamma)=\inf_{N}\left[ \alpha N+ \mathcal{G}(N,\beta,\gamma)\right].
\label{rescaled_replica_LFT_3}
\end{equation}
Now let us consider the response function $M_{T,P}$ in the isothermal-isobaric ensemble.
The rescaled Gibbs free energy is not necessarily convex in $N$ for nonadditive systems, so that
\begin{equation}
\frac{1}{M_{T,P}}=\left(\frac{\partial \mu}{\partial N}\right)_{T,P}= \left(\frac{\partial^2 G}{\partial N^2}\right)_{T,P}
\end{equation}
is not restricted to be a positive quantity. The unconstrained and isothermal-isobaric ensembles are not equivalent if the Gibbs free energy
does not coincide with its convex envelope with respect to $N$.
The response function $M_{T,P}$ in the isothermal-isobaric ensemble can be negative at points where the rescaled Gibbs free energy is not locally convex
with respect to $N$, or could have points where it is not defined, or both (the three cases of Figure~\ref{fig_concave}).

In analogy with the previous situations, here we point out that if the rescaled Gibbs free energy
energy does not coincide with its convex envelope with respect to $N$, the isothermal-isobaric and the unconstrained ensembles are
not equivalent. In this case, the function $\mathcal{R}$ presents
at least a point of discontinuous derivative with respect to $\alpha$, associated to a first order phase transition.
If an equilibrium isothermal-isobaric state in which the rescaled Gibbs free energy does not coincide with its convex envelope with respect to $N$
is put in contact with an environment with its same chemical potential and with which it can exchange particles, then it becomes either unstable or
not globally stable.

A final remark. We have noted that the replica energy $\mathscr{E}$ vanishes for additive systems. This is related to the
fact that the validity of the Gibbs-Duhem equation implies that the variables $(T,P,\mu)$ cannot be taken as independent control
parameters for those systems. In turn, this implies that $\Upsilon = \e^{-\mathcal{R}}$ is negligible in the thermodynamic limit.

\section{Discussion}
\label{discuss}

\begin{figure}[h]
\centering
\includegraphics[scale=0.7,trim= 0cm 22cm 0cm 2cm,clip]{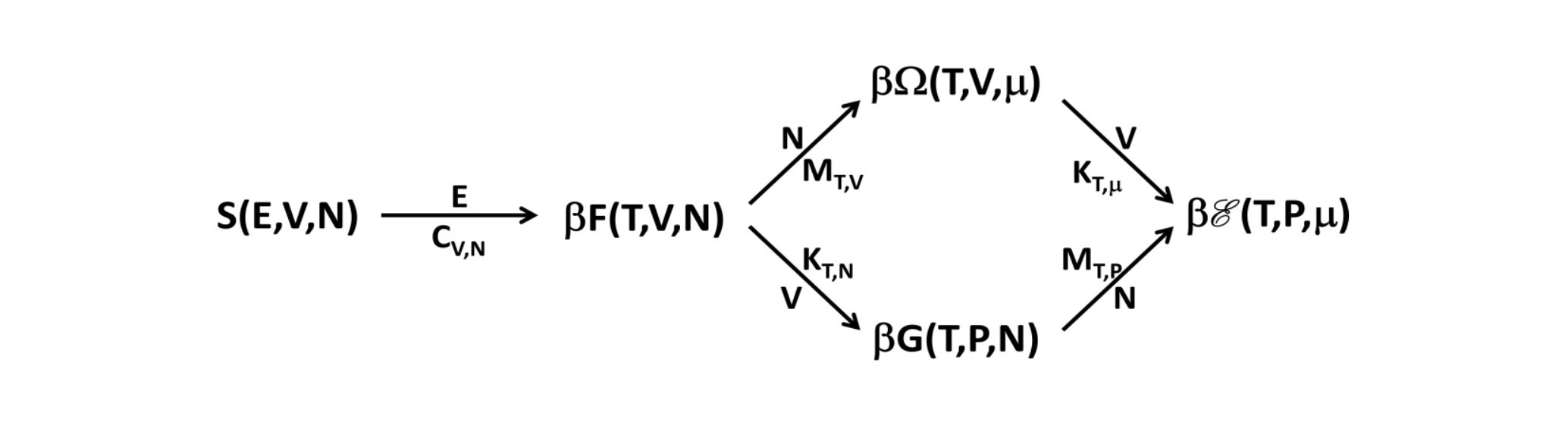} 
\caption{A schematic picture showing the connection between the thermodynamic functions through Legendre-Fenchel transformations. For each
thermodynamic function, the natural variables on which it depends are shown. The arrows connect the starting and the arriving function
of the Legendre-Fenchel transformations. On one side of each arrow there is the constraint variable with respect to which one has to minimize
to perform the transformation; on the other side of the arrow there is the response function associated with the possible
ensemble inequivalence: the response function is always positive in the arriving thermodynamic function, while it can be negative in the
starting function if ensemble inequivalence occurs. Each response function concerns the response of the constraint variable of the corresponding
Legendre-Fenchel transformation with respect to its conjugate thermodynamic variable, while keeping constant the other
two variables (shown in the subscripts) on which the arriving functions depend on. Actually, except in the first transformation, relating $S$
and $\beta F$, the starting function of the other Legendre-Fenchel transformations are given by minus the indicated function (see text). However,
this is irrelevant for our general discussion.}
\label{fig_scheme}
\end{figure}

We have seen that ensemble inequivalence is connected with the occurrence of negative response functions, and that these
anomalous responses are in turn associated to anomalous concavity properties of the thermodynamic functions. In details,
we note that all these response functions concern the variation of a constraint variable ($E$, $V$ or $N$) with respect to the respective
conjugate thermodynamic variable ($T$, $P$ and $\mu$, respectively). Also, the Legendre-Fenchel transformations relating the various
thermodynamic functions are defined by the minimization with respect one of the constraint variables. In Figure~\ref{fig_scheme}, we show
a simple scheme of the transformations and of the response functions connecting the different thermodynamic potentials.

It is interesting to note the following. The rescaled replica energy $\beta \mathscr{E}$ is obtained from the microcanonical entropy by minimizing
with respect to all the constraint variables. However, while the first minimization with respect to $E$ produces the rescaled Helmholtz free energy $\beta F$, the following minimizations with respect to $N$ and $V$ can be made in the two different orders, thus producing, as
``intermediate'' thermodynamic functions, either $\beta \Omega$ or $\beta G$. For this reason the scheme in Figure~\ref{fig_scheme} has
two routes from $S$ to $\beta \mathscr{E}$.

We stress once more that a negative response function implies ensemble inequivalence, while the reverse is not true: ensemble inequivalence
can occur with or without a negative response function. We have described in each case which response function can be negative, and referring to Figure~\ref{fig_concave}
we have cited the possible situations, clarifying that ensemble inequivalence implies the presence of a first order phase transition in the
ensemble which is the arriving one in the Legendre-Fenchel transformation. We also note that, very often, concrete models can present all the three cases
considered in Figure~\ref{fig_concave}, that occur varying the value of the parameters of the Hamiltonian.

In principle one may wonder about the following point. Is it possible that an ensemble corresponding to a thermodynamic function that is the
starting one in a Legendre-Fenchel transformation is not equivalent to the ensemble corresponding to the arriving function, but at the same time
it is equivalent to the ensemble corresponding to a successive function of the scheme? For more clarity and as an example, referring to
Figure 1: is it possible that the microcanonical ensemble, corresponding to $S$, is not equivalent to the canonical ensemble, corresponding to
$\beta F$, but it is equivalent either to the grand canonical ensemble, corresponding to $\beta \Omega$, or the isothermal-isobaric ensemble,
corresponding to $\beta G$? This situation is not possible and it can be seen in the following way. Suppose that the microcanonical ensemble is
equivalent to the grand canonical ensemble. This means that Eq. (\ref{rescaled_Omega_LFT_2}) can be inverted, obtaining
\begin{equation}
S(E,V,N)=\inf_{\alpha,\beta}\left[\alpha N+\beta E-\mathcal{L}(\alpha,\beta,V)\right] \, ,
\label{micro_from_grand}
\end{equation}
i.e., the microcanonical entropy $S(E,V,N)$ coincides with its concave envelope with respect to the double Legendre-Fenchel transformation in the
$(E,N)$ plane. This implies
that $S(E,V,N)$ is globally concave in the $(E,N)$ plane; but then it is a fortiori globally concave with respect to $E$, and it coincides with
its concave envelope with respect to $E$. In turn, this implies that the microcanonical and the canonical ensembles are equivalent. Then, if
the microcanonical ensemble is not equivalent to the canonical ensemble, it is not equivalent also to the grand canonical ensemble.

The same procedure can be used if one assumes that the microcanonical ensemble is equivalent to the isothermal-isobaric ensemble. Then
Eq. (\ref{rescaled_G_LFT_2}) can be inverted to have
\begin{equation}
S(E,V,N)=\inf_{\beta,\gamma}\left[\beta E+\gamma V-\mathcal{G}(N,\beta,\gamma)\right] \, ,
\end{equation}
i.e., the microcanonical entropy coincides with its concave envelope with respect to the double Legendre-Fenchel transformation in the $(E,V)$ plane.
This implies that $S(E,V,N)$ is globally concave in the $(E,V)$ plane; but then, as before, it is a fortiori globally concave with respect to $E$ and the microcanonical and canonical ensembles are equivalent. Then, if the microcanonical ensemble is not equivalent to the canonical ensemble, it is not equivalent also to the isothermnal-isobaric ensemble.

The above derivations are valid regardless of the differentiability of the thermodynamic functions. It is instructive to give also an
alternative derivation based on partial derivatives, that shows that the heat capacity at constant $V$ and $N$ is positive in both the
grand canonical and isothermal-isobaric ensemble. This is not completely trivial, since $(\beta,V,N)$ are not the control parameters
of either of these two ensembles. In Appendix 2 we present this derivation.

In an analogous way, if the canonical ensemble is not equivalent to, e.g., the grand canonical ensemble, then it is not equivalent also to the
unconstrained ensemble. On the contrary, it may happen that canonical and grand canonical ensembles are equivalent, but they are both not equivalent to the unconstrained ensemble (see Ref.~\cite{Latella_2017} for a concrete example).

In this paper we have presented a general discussion, without reference to any specific model. Although the results are valid
regardless of the differentiability of the thermodynamic functions, as a matter of fact, the most interesting situations arise when we have points
where the differentiability does not hold, i.e., when we are dealing
with first-order phase transitions. In fact, if neither of the two ensembles connected by a Legendre-Fenchel transformation
has a first order phase transition, but at most a continuous transition, then the two ensembles are equivalent.

The results here discussed have the consequence that with ensemble inequivalence the phase
transitions are located, generally, in different points of the thermodynamic phase diagram for nonadditive systems. From the general
results one can also prove that, in many cases, it is possible to obtain the response function in the ``higher'' ensemble from that
in the ``lower'' ensemble (where ``higher'' means that, in the scheme of Figure 1, it is on the right of the ``smaller'' and
connected by one or more arrows) by invoking the Maxwell construction. For example, if one computes the function $T(E)$ (at constant $V$ and
$N$) in the microcanonical ensemble and then obtain the specific heat $C_{V,N}$, then the analogous curve and the specific heat
in the canonical ensemble are obtained by applying the Maxwell construction in the neighborhood of the regions where the microcanonical
$C_{V,N}$ is negative.

In any case, we believe that a general and simple scheme like the one given in this paper can be useful as a reference material when dealing
with concrete nonadditive systems.

\section*{Acknowledgments}
A. C. acknowledges financial support from INFN (Istituto Nazionale di Fisica Nucleare) through the projects DYNSYSMATH and ENESMA.

\section*{Appendix 1: Legendre-Fenchel transformation, concave and convex functions}

In this appendix we recall, without proof, some relations between concave functions and the Legendre-Fenchel transformation.
A function $f(x)$ is said to be concave if the relation
\begin{equation}
f\left(cx_1 + (1-c)x_2\right) \ge cf(x_1) + (1-c)f(x_2)
\label{eq_app1}
\end{equation}
holds for any $0\le c \le 1$. If this occurs for any $(x_1,x_2)$ in the range of definition of $f$, the function is said to be
globally concave; if the relation is satisfied only for $x_1$ and $x_2$ belonging to a sufficiently small neighborhood of a point $x$,
then $f$ is said to be locally concave in $x$. From the practical point of view, the graph of $f$ between $x_1$ and $x_2$ lies above
the straight line connecting $f(x_1)$ and $f(x_2)$. If $f$ is twice differentiable, then its second derivative is nonpositive in a point
of local concavity and it is nonpositive in the whole range of definition for a globally concave function.

\begin{figure}[h]
\centering
\includegraphics[scale=0.6,trim= 0cm 9cm 0cm 6cm,clip]{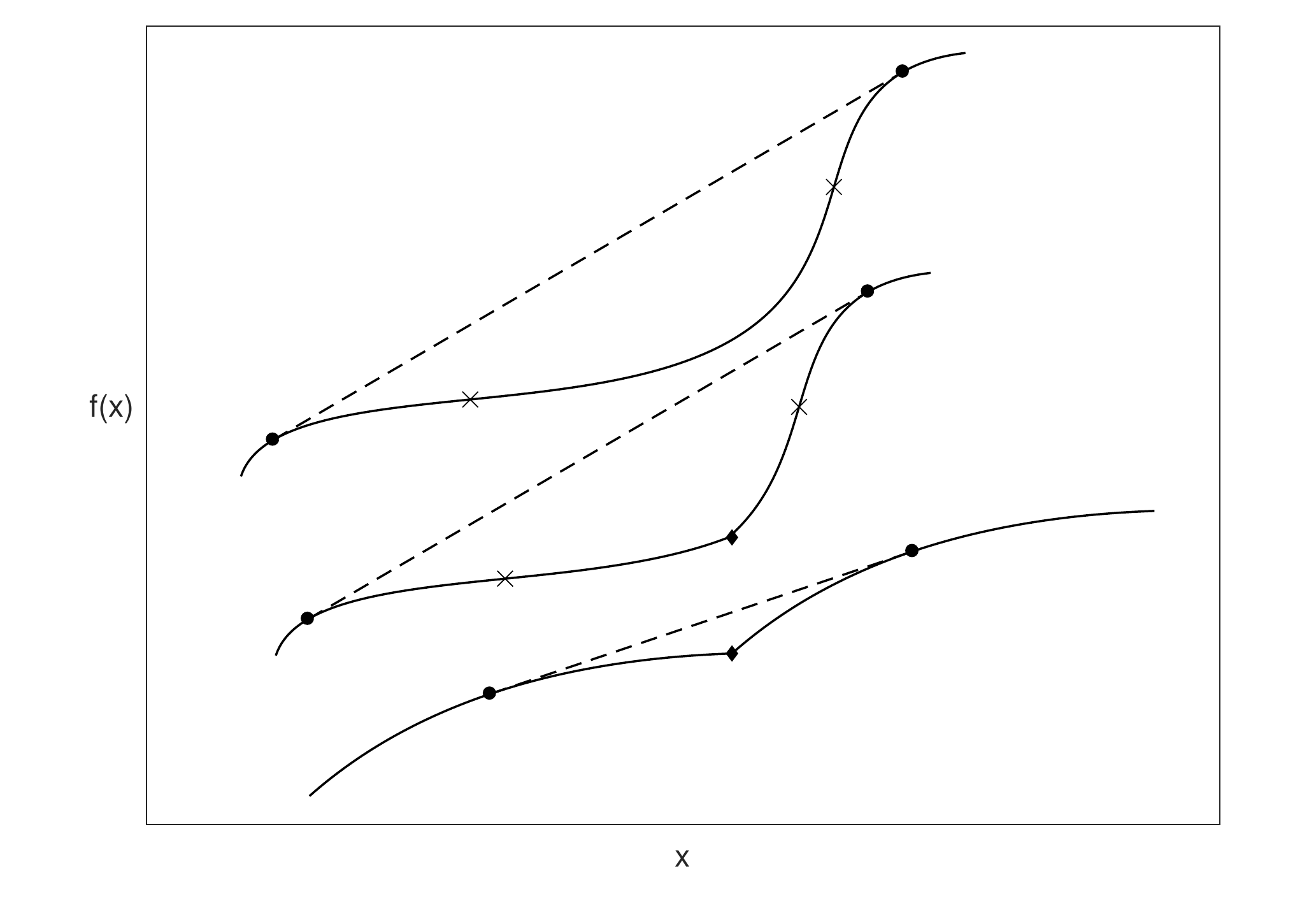} 
\caption{A representative plot showing three situations often occurring in the study of nonadditive systems. We can think of $x$ as a constraint variable,
e.g., $E$, and $f$ as a thermodynamic function, e.g., $S$. The three curves in solid lines show functions that are not globally concave; the upper curve is
twice differentiable, while the other two curves have a discontinuous derivative at the point of the cusp, marked by a diamond. The upper and the middle curves
are locally concave outside the $x$ range between the two crosses, where their second
derivative is negative, while in the range between the crosses their second derivative is positive. On the contrary, the lower curve is locally concave everywhere
except at the point of discontinuity of its first derivative, since the second derivative is always negative except at that point.
The dashed lines define the concave envelope of each function; more precisely, the concave envelope $f^{**}(x)$ is equal to the dashed line in the $x$ range where
this line is defined, while it is equal to $f(x)$ outside this range. Note that the range where $f^{**}(x)$ does not coincide with
$f(x)$ is larger than that where the function is not locally concave. The dots marking the ends of the dashed lines are just for visual clarity.}
\label{fig_concave}
\end{figure}

The Legendre-Fenchel transform $g(z)$ of $f(x)$ is defined by
\begin{equation}
g(z) = \inf_x \left[ zx - f(x) \right] \, .
\label{eq_app2}
\end{equation}
It is easy to show that $g(z)$ is globally concave. By applying the Legendre-Fenchel transformation to $g(z)$ (loosely speaking,
by inverting the transformation) we obtain the following function of $x$:
\begin{equation}
f^{**}(x) = \inf_z \left[ xz - g(z) \right] \, ,
\label{eq_app3}
\end{equation}
where we have adopted a common notation for functions obtained by applying twice the transformation. The function $f^{**}(x)$ is called
the concave envelope of $f(x)$. Being defined by a Legendre-Fenchel transformation, $f^{**}(x)$ is globally concave. If the starting function $f(x)$ is
globally concave, then $f^{**}(x)$ coincides with it; otherwise it is the smallest globally concave function which is larger than $f(x)$
(where $f_1(x$) smaller than $f_2(x)$ here means that $f_1(x) < f_2(x)$ for any $x$ in the range of definition). In Figure~\ref{fig_concave}, we show
three examples of functions that are not globally concave and thus do not coincide with their concave envelope. In the upper curve the function is twice
differentiable, while in the other two curves the first derivative has a point of discontinuity. Apart from the point of discontinuity, the lower curve
has always a negative second derivative (i.e., it is locally concave), while the middle curve, like the upper one, has a range where the second derivative
is positive. In should be noted that
the range where $f^{**}(x)$ and $f(x)$ do not coincide is larger than the range where $f(x)$ is not locally concave.
The following important result can be proved. Whenever the concave envelope $f^{**}(x)$ does not coincide with $f(x)$ (e.g., in all cases represented
in Figure~\ref{fig_concave}) the Legendre-Fenchel transform $g(z)$ has at least a point where its first derivative is discontinuous. Equivalently, if $g(z)$
is differentiable, in particular if furthermore it is twice differentiable, then the concave envelope $f^{**}(x)$ coincides with $f(x)$,
i.e., $f(x)$ is globally concave.

Let us now consider a function of two variables, $f(x,y)$. It can be concave (locally or globally) as a function of $x$ for a given $y$,
and/or as a function of $y$ for a given $x$. If it is concave in both variables and twice differentiable, we have (adopting the usual notation for derivatives)
$f_{xx}<0$ and $f_{yy}<0$. Concavity with respect to each of the two variables is necessary, but not sufficient to make $f$ completely concave
[or in other words concave in the plane $(x,y)$]; complete concavity is defined by 
\begin{equation}
f\left(cx_1 + (1-c)x_2,cy_1 + (1-c)y_2\right) \ge cf(x_1,y_1) + (1-c)f(x_2,y_2)
\label{eq_app4}
\end{equation}
for any $0\le c \le 1$.
For twice differentiable functions, to have complete concavity we must have also
$f_{xx}f_{yy} - f_{xy}^2 >0$. The Legendre-Fenchel transform
with respect to $x$
\begin{equation}
g(z,y) = \inf_x \left[ zx - f(x,y) \right]
\label{eq_app5}
\end{equation}
is globally concave with respect to $z$, while the Legendre-Fenchel tranform with respect to $y$
\begin{equation}
g(x,w) = \inf_y \left[ wy - f(x,y) \right] 
\label{eq_app6}
\end{equation}
is globally concave with respect to $w$. On the other hand, the double Legendre-Fenchel transform
\begin{equation}
g(z,w) = \inf_{x,y} \left[ zx + wy - f(x,y) \right] 
\label{eq_app7}
\end{equation}
is globally and completely concave in the $(z,w)$ plane. These definitions and properties for functions of two variables can be readily
extended, with obvious modifications, to functions of more than two variables.

The above properties have the analogous ones by defining a transformation similar to (\ref{eq_app2}), but with a supremum instead
of an infimum, i.e.,
\begin{equation}
g_c(z) = \sup_x \left[ zx - f(x) \right] .
\label{eq_app8}
\end{equation}
Let us call this transformation, just to distinguish it from the previous one, the convex Legendre-Fenchel transformation; accordingly, we have
put a subscript to $g(z)$.
We can also define convex functions which satisfy an inequality similar to (\ref{eq_app1}), but where the left hand side
is smaller than or equal to the righ hand side [the graph of $f$ between $x_1$ and $x_2$ lies below
the straight line connecting $f(x_1)$ and $f(x_2)$], namely,
\begin{equation}
f\left(cx_1 + (1-c)x_2\right) \le cf(x_1) + (1-c)f(x_2).
\label{eq_app9}
\end{equation}
Then it can be proved that the convex Legendre-Fenchel transformation gives rise to globally convex
functions~\cite{Touchette_2009}. In analogy with the previous case, we can define the convex envelope of $f(x)$ by
\begin{equation}
f^{**}_c(x) = \sup_z \left[ xz - g_c(z) \right] .
\label{eq_app10}
\end{equation}
It follows that if $f(x)$ is globally convex, then $f^{**}_c(x)$ and $f(x)$ coincide.

In the main text we find Legendre-Fenchel transformations in which the transformed function is not $f(x)$, but $-f(x)$, as in
\begin{equation}
g(z) = \inf_x \left[ zx + f(x) \right] \, .
\label{eq_app11}
\end{equation}
This can also be written as
\begin{equation}
-g(z) = \sup_x \left[ -zx - f(x) \right] \, .
\label{eq_app12}
\end{equation}
Thus, $-g(z)$ is the convex Legendre-Fenchel transform of $f(x)$ (the fact that there appears $-zx$ instead of $zx$ is irrelevant),
namely, $g(z)=-g_c(z)$.
If $-f(x)$ coincides with its concave envelope, that is, if
\begin{equation}
-f(x) = \inf_z \left[xz - g(z) \right] \, ,
\label{eq_app13}
\end{equation}
then $f(x)$ coincides with its convex envelope, since the last expression can also be written as
\begin{equation}
f(x)= \sup_z \left[ -xz - g_c(z) \right].
\label{eq_app14}
\end{equation}
This conclusion can be easily inferred in a visual manner. In fact, it is trivial to see, e.g., that if $f(x)$ is twice differentiable and its second derivative is negative (positive) definite, then the second derivative of $-f(x)$ is positive (negative) definite.
In the main text we thus refer to global convexity for thermodynamic functions $f(x)$ where the Legendre-Fenchel transformation
involves $-f(x)$.

\section*{Appendix 2: Heat capacity in the grand canonical and the isothermal-isobaric ensembles}

Here we show that the response function $C_{V,N}$ is positive not only in the canonical ensemble, but also in the grand canonical ensemble
and in the isothermal-isobaric ensemble. In this Appendix we assume that the thermodynamic functions are twice differentiable.
Thus, we show that if the microcanonical ensemble is not equivalent to the canonical ensemble, i.e., if $C_{V,N}$
is negative in the microcanonical ensemble, then this ensemble is not equivalent also to the grand canonical and the isothermal-isobaric ensembles.
Summarizing, we want to see that the quantity
\begin{equation}
\label{for_example}
\left( \frac{\partial \bar{E}}{\partial \beta}\right)_{V,N} =-\beta^2C_{V,N}
\end{equation}
is negative in both the grand canonical and the isothermal-isobaric ensembles. This is not completely trivial, since $(\beta,V,N)$ are not the
control parameters of either of the two ensembles. We proceed as follows. Beginning with the grand canonical ensemble, we start from the relations
\begin{eqnarray}
\label{egrand}
\bar{E} &=& \left( \frac{\partial (\beta \Omega)}{\partial \beta}\right)_{V,\beta \mu} \\
\label{Ngrand}
\bar{N} &=& -\left( \frac{\partial (\beta \Omega)}{\partial (\beta \mu )}\right)_{\beta,V} \, .
\end{eqnarray}
Then, we have to compute the second partial derivative of $\beta \Omega$ with respect to $\beta$ while keeping constant its first
partial derivative with respect to $(\beta \mu)$. Without showing the passages, we state the result. One obtains
\begin{eqnarray}
\label{cvngrand}
\left( \frac{\partial \bar{E}}{\partial \beta}\right)_{V,N} &=& 
\left[ \left( \frac{\partial^2 (\beta \Omega)}{\partial (\beta \mu)^2}\right)_{\beta,V}\right]^{-1} \nonumber \\
&\times&\left[ \left( \frac{\partial^2 (\beta \Omega)}{\partial \beta^2}\right)_{\beta \mu,V} \left( \frac{\partial^2 (\beta \Omega)}
{\partial (\beta \mu)^2}\right)_{\beta,V} -\left(\frac{\partial^2 (\beta \Omega)}{\partial \beta \partial (\beta \mu)}\right)_{V}^2\right] \, .
\end{eqnarray}
We have noted in Section \ref{ensemble_inequivalence} that $\beta \Omega$ is completely concave in the $(\beta \mu, \beta)$ plane. This implies
that the quantity in square brackets in the second line of the expression is positive, while that in the first line is negative. Therefore, we
obtain a negative quantity. For the isothermal-isobaric ensemble, we can proceed in an analogous way. Now we start from the relations
\begin{eqnarray}
\label{eisob}
\bar{E} &=& \left( \frac{\partial (\beta G)}{\partial \beta}\right)_{N,\beta P} \\
\label{Visob}
\bar{V} &=& \left( \frac{\partial (\beta G)}{\partial (\beta P )}\right)_{\beta,N} \, .
\end{eqnarray}
Then, we have to compute the second partial derivative of $\beta G$ with respect to $\beta$ while keeping constant its first
partial derivative with respect to $(\beta P)$. The result is now
\begin{eqnarray}
\label{cvisob}
\left( \frac{\partial \bar{E}}{\partial \beta}\right)_{V,N} &=& 
\left[ \left( \frac{\partial^2 (\beta G)}{\partial (\beta P)^2}\right)_{\beta,N}\right]^{-1} \nonumber \\*
&\times&\left[ \left( \frac{\partial^2 (\beta G)}{\partial \beta^2}\right)_{\beta P,N} \left( \frac{\partial^2 (\beta G)}
{\partial (\beta P)^2}\right)_{\beta,N} -\left(\frac{\partial^2 (\beta G)}{\partial \beta \partial (\beta P)}\right)_{N}^2\right] \, .
\end{eqnarray}
We have also noted in Section \ref{ensemble_inequivalence} that $\beta G$ is completely concave in the $(\beta P, \beta)$ plane. This implies
that the quantity in square brackets in the second line of the expression is positive, while that in the first line is negative. Therefore we
obtain again a negative quantity.

\vspace{6pt} 

\section*{References}

\end{document}